# Orientation-Resolved Nanoindentation of Single Crystal Ti-15Mo: Inverted Elastic Anisotropy and Yield-Onset Behavior Beyond Simple CRSS Predictions

K. Abebe[a], D. Gaetano[a], A. Moriyama-Gurish[a], J. Šmilauerová[b], Y.F. Gao[c], U.D. Schwarz[a,d], A. Datye[a]

[a]Department of Mechanical Engineering and Materials Science, Yale University, New Haven, CT, 06511
[b]Department of Physics of Materials, Charles University, Prague, Czech Republic
[c]Department of Materials Science and Engineering, University of Tennessee, Knoxville, TN, 37996
[d]Department of Chemical and Environmental Engineering, Yale University, New Haven, CT, 06511

## Abstract

Ti-15Mo (wt%) is a well-established β-stabilized alloy whose mechanical response is strongly governed by phase stability and deformation mechanisms that operate at the single-grain scale. Because these intragranular mechanisms are intrinsically anisotropic, understanding the orientation dependence of the elastic and plastic response is critical to the texture-engineered processing of wrought components. Here, we quantify the nanomechanical anisotropy of single-crystal Ti–15Mo by combining electron backscatter diffraction (EBSD) with nanoindentation to correlate hardness, indentation modulus, strain rate sensitivity, and indentation yield response with the crystallographic surface normal near ⟨111⟩, ⟨124⟩, ⟨011⟩, ⟨023⟩, and ⟨013⟩ directions.

While the modulus anisotropy is small (~10%), contrary to the prediction for a cubic crystal with A > 1, orientations near ⟨111⟩ exhibit the lowest moduli. Yield onset is likewise softest near ⟨111⟩ and spans 41%, while hardness spans only 6.6% and follows neither the modulus nor the yield ordering. The alloy also exhibits yield behavior not captured by the simple CRSS/Schmid-factor models evaluated here. Strain-rate-jump tests yield $V^* = 18–31\ b^3$, consistent with kink-pair-nucleation-controlled plasticity; based on V*, the orientations separate into two groups suggestive of glide-dominated versus twin/transformation-participating response.

We reveal an inversion of the elastic anisotropy expected for a homogeneous cubic β-phase and an orientation-dependent yield-onset response that is not reproduced by the simple Schmid-factor models examined here. The observations are consistent with an additional orientation-sensitive anelastic contribution to the indentation response. Retained athermal ω is proposed as the contributor because β→ω collapse occurs on {111}β planes and may provide anelastic compliance through the reversible ω→β transformation. However, the present measurements don't establish the ω-phase fraction or variant population. Site-specific diffraction/TEM measurements are therefore required to test this mechanism. The strain-rate-jump measurements show resolvable orientation dependence in the apparent activation volume compatible with thermally activated plasticity but don't uniquely isolate the controlling mechanism.

## 1 Introduction

Titanium (Ti) –based alloys are widely used in biomedical and aerospace systems due to their exceptional strength-to-weight ratio, fatigue life, and corrosion resistance [1, 2]. In load-bearing implants, the ability to reduce elastic modulus while maintaining high strength is especially important for mitigating stress shielding and improving long-term fixation. Metastable β-titanium alloys have therefore attracted sustained attention, since β-stabilizing elements can retain the body-centered cubic (BCC) phase at room temperature and enable comparatively low moduli relative to α-phase titanium [3, 4]. Among these materials, Ti with 15 wt% Molybdenum (Mo) (ASTM F2066 [5]) combines high strength with reduced modulus and excellent corrosion resistance, while also avoiding concerns associated with the cytotoxicity of Aluminum (Al) / Vanadium (V) –containing alloys such as Ti–6Al–4V [6-8].

Prior studies on Ti–15Mo report tensile strengths approaching ~900 MPa with moduli on the order of ~80–110 GPa depending on processing route and microstructure, highlighting the alloy's promise for structural and biomedical applications [7, 9, 10]. The mechanical response of Ti-15Mo is governed by a combination of phase stability and deformation mechanisms. Molybdenum stabilizes the β phase at room temperature [4], while secondary phases such as α′ or ω can precipitate depending on cooling rate and aging, substantially altering hardness and modulus [7, 11, 12]. For example, aging treatments have been shown to change Ti–15Mo hardness and stiffness in ways consistent with precipitation and dissolution of nanoscale ω, emphasizing that nanoscale phase constitution plays a critical role in strengthening and compliance [13]. Metastable titanium-based alloys have also shown to possess strength and ductility enhancing deformation mechanisms, ranging from twinning-induced plasticity (TWIP) to stress-induced phase transformations [14]. Current Ti-15Mo literature reports predominantly polycrystalline, orientation-averaged data. Because each of the mechanisms above is tied to specific crystallographic planes and directions, orientation-averaged measurements cannot resolve which mechanism operates in a given grain, motivating an orientation-resolved study.

Orientation-specific mechanics of Ti-15Mo are largely unexamined, but the broader β-Ti literature offers guidance: numerous single-crystal studies have probed the anisotropic elastic and plastic response of these alloys. First-principles surveys of Ti–X (X = V, Nb, Ta, etc.) alloys show that the Zener anisotropy ratio traces a U-shaped dependence on valence-electron concentration and passes through isotropy only near $e/a \approx 4.7$. This places binary β alloys that stabilize near $e/a \approx 4.2$, such as Ti-15Mo, deep in the anisotropic regime [15, 16]. Uniaxial Young's modulus measurements of single crystal β-Ti alloys corroborate these computational studies: $E\langle 100\rangle \approx 35$ GPa vs $E\langle 111\rangle \approx 70$ GPa in Ti–29Nb–13Ta–4.6Zr [17], and $E\langle 100\rangle \approx 44.4$ GPa vs $E\langle 111\rangle \approx 120$

GPa in Ti–15Mo–5Zr–3Al [18]. Orientation governs plastic flow as well: the elastic and plastic anisotropies in β-Ti alloys have been shown to be empirically coupled, with grains oriented along the compliant orientations also being those in which slip activity and strain localization first appear [19, 20]. Furthermore, studies on micropillar compression attribute orientation-dependent flow in β-Ti alloys to interactions between ⟨111⟩ slip and the four ω-phase variants [21, 22]. On the nanoindentation scale, high-speed indentation mapping of β-Ti has reported no measurable orientation dependence of hardness or modulus [23], and strain rate sensitivity in single-phase β has likewise been found orientation-insensitive [24].

Despite the extensive work on bulk mechanical behavior, microstructure evolution, and processing of Ti–15Mo in polycrystalline and additively manufactured forms, and the broader literature focused on mechanical anisotropy in metastable β-Ti alloys, the interaction of the indentation stress state, orientation dependent local deformation mechanisms and nanoscale ω-phase related structures, is not well understood. This matters because orientation-dependent behavior underpins the processing and design of texture-engineered and single-crystal components. Here we quantify the mechanical anisotropy of single-crystal Ti-15Mo, focusing on the elastic modulus, hardness, strain rate sensitivity, and yield strength along the ⟨111⟩, ⟨124⟩, ⟨011⟩, ⟨023⟩, and ⟨013⟩ directions, by combining EBSD with nanoindentation. We find that these properties vary with orientation at different magnitudes, in different rank orderings, and at different strain scales, and the observed trends cannot be rationalized by homogeneous β-phase elasticity and simple resolved-shear-stress arguments. Because these traditional frameworks do not consistently reproduce the observed trends, we propose that mechanisms emanating from retained athermal and deformation-induced ω-phase as a working hypothesis, recognizing direct confirmation requires nanoscale microstructural analysis or in-situ experimentation.

## 2 Materials and Methods

### 2.1 Sample Preparation

The sample is a single-crystal BCC Ti-15Mo (wt%) alloy [25] in the form of a cylinder approximately 6 mm in diameter and 25 mm in length. The sample was manufactured through directional growth and solidification using a seed crystal as an orientation template. The sample was annealed at 860 ºC for 4 hours and water-quenched. The cylinder was sliced into disks that were 2 mm thick. Using these disks in conjunction with EBSD, several smaller samples were sliced at various angles to expose different crystallographic orientations. A beta-annealed polycrystalline Ti-15Mo sample (Goodfellow Advanced Materials, Part #1000171649) and was also sliced into 2 mm thick discs for mounting and polishing. All the samples were mounted in resin and mechanically polished to nanoindentation and EBSD quality (< 1 nm root-mean-squared roughness, verified through atomic force microscopy (AFM) using the Bruker Dimensions Fastscan AFM).

## 2.2 SEM and EBSD

A Hitachi SU 7000 scanning electron microscope (SEM) and the Oxford Instruments Symmetry S3 detector, were used for EBSD measurements. An accelerating voltage of 20 kV was used, and SEM images were taken of the polycrystalline sample to estimate the mean grain diameter to ensure that our mechanical tests represent an orientation-averaged response. To perform the EBSD measurements, the sample was tilted at 69º. In the AZtecHKL software, the "Point ID" tool was used to obtain the Euler angles at a point on the sample surface, which were then converted to the Miller indices of each orientation.

## 2.3 Mechanical Characterization

<u>CSM hardness and modulus measurements</u>

Nanoindentation was performed using an iNano nanoindenter (KLA, Milpitas CA) with a Berkovich indenter (Micro Star Technologies, Huntsville, TX) and possessing a modulus and Poisson's ratio of 1140 GPa and 0.07, respectively) for basic hardness and modulus measurements and strain rate jump (SRJ) tests. Nanoindentation was also performed with a spherical indenter (nominal radius = 2.5 µm, Micro Star Technologies) to confirm azimuthally-independent modulus trends. A modified continuous stiffness measurement (CSM) method [26, 27] with improved surface contact detection based on the Oliver-Pharr method [28, 29] was used to make at least 16 indents in each sample at a strain rate of 0.2 /s up to 50 mN with a grid spacing of 50 µm. This method was implemented with both the Berkovich and Spherical indenters. The Bruker Dimensions Fastscan AFM, with the Fastscan-B tips (nominal radius = 5 nm, Bruker Corp., Camarillo, CA) and the Nanoscope Analysis 3.00 software were used to characterize the material pile-up around the CSM indents. Based on the method outlined by Qui, Y. H. et al. [30], the hardness and reduced modulus measurements were corrected for the pile-up effect.

<u>Strain rate jump tests</u>

The SRJ tests were performed to a peak load of 50 mN over a strain rate range of 0.005/s to 0.5/s, and the indent markers were placed in a 4x4 grid, each marker spaced 50 µm apart, making a total of 16 indents.

<u>Flat punch tests</u>

Finally, using the 2.25 µm diameter flat punch indenter (Micro Star Technologies), nanoindentation was performed to a peak load of 40 mN at a strain rate of 0.001/s for a total of 16 indents. To obtain the equivalent stress-strain curves and yield properties of each sample, we adopt the method developed and patented by KLA Instruments [26, 31]. As part of the analysis of the flat punch data, we compute the indentation Schmid factors for all the single crystal samples adapting the method outlined by Li et al. 2011 [32], detailed in Appendix B.

## 3 Results

### 3.1 EBSD

EBSD measurements confirmed the orientations of all five single crystal samples. Table 1 below summarizes the Euler angles and corresponding orientations, and Figure 1 shows the EBSD map of a representative sample and an inverse pole figure (IPF) plot of all the orientations in this study. Note that the rounded miller indices we adopt exhibit low misorientations (< 6°), which is significantly better than accepted misorientation tolerances in metallurgical texture research [33].

Table 1: Bunge convention Euler angles ($\phi_1, \Phi, \phi_2$) and the nearest corresponding orientation for each sample. We also illustrate the approximate misorientation from the true orientation axis.

| $\phi_1$ (°) | $\Phi$ (°) | $\phi_2$ (°) | Approximate Orientation | Misorientation (°) |
|---|---|---|---|---|
| 22.1 | 127.8 | 224.2 | ⟨111⟩ | 2.61 |
| 347.1 | 150.6 | 198.2 | ⟨124⟩ | 4.09 |
| 81 | 135.4 | 264.6 | ⟨011⟩ | 3.82 |
| 334.09 | 34.72 | 81.72 | ⟨023⟩ | 4.76 |
| 268.51 | 19.67 | 17.82 | ⟨013⟩ | 5.92 |

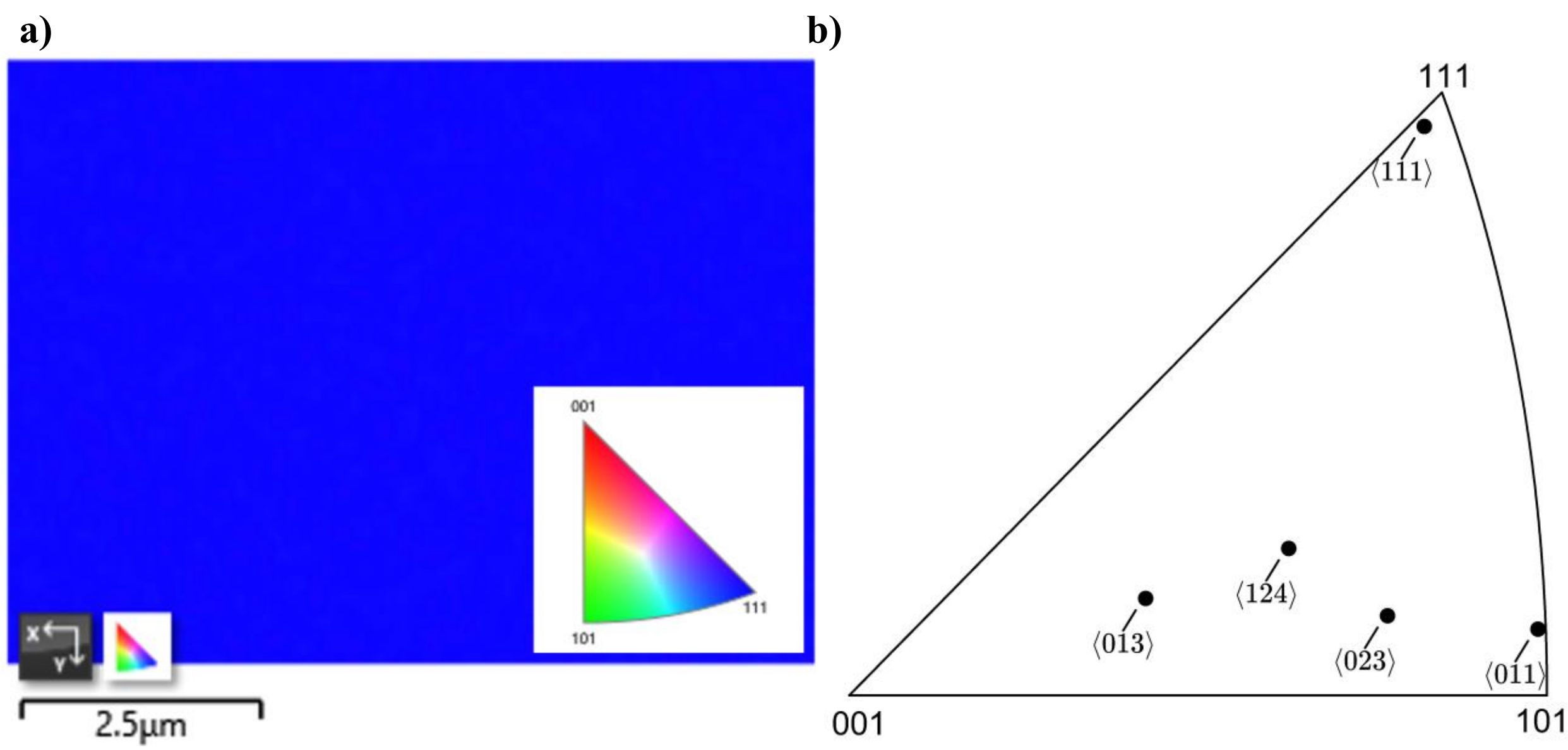

Figure 1: **a)** The EBSD map and corresponding IPF figure of the ⟨111⟩ sample. This map exhibits a uniform phase because the athermal ω developed as a byproduct of the solution treatment process is on the nanometer scale, smaller than the resolution capabilities of the SEM/EBSD used here. **b)** an IPF figure showing all the single crystal orientations.

3.2 Modulus and Hardness

The elastic constants of identically processed Ti-15Mo alloy from literature, ($C_{11} = 139.4 \pm 1$ GPa, $C_{12} = 95.7 \pm 0.6$ GPa, and $C_{44} = 44.7 \pm 0.6$ GPa) [25], were used to calculate the predicted anisotropic modulus ($M_{\text{predicted}}$) using the Vlassak/Nix formalism [34-36], here after referred to as V/N and detailed in Appendix A. The in-plane indentation modulus ($M_{\text{corrected}}$) was obtained after pile-up correction (see Appendix C) and compared to $M_{\text{predicted}}$. The central result is an inverted rank order of the modulus ($M_{\text{corrected}}$): $\langle 111 \rangle$, predicted to be the stiffest direction at ~108 GPa, is the softest at 85.40 ± 0.94 GPa, while $\langle 013 \rangle$, predicted the most compliant at ~103 GPa, measures among the stiffest at 91.71 ± 1.39 GPa. The polycrystal has a Young's modulus (E) of 76.07 ± 1.98 GPa (calculated using the Oliver-Pharr equation and a Poisson's ratio of 0.3, see Appendix A) consistent with reported values for β-phase polycrystalline Ti–15Mo (~70 – 80 GPa) [9, 37]. The pile-up-corrected hardnesses span 3.04 ± 0.07 GPa ($\langle 111 \rangle$) to 3.24 ± 0.06 GPa ($\langle 023 \rangle$), with the polycrystal at 3.07 ± 0.12 GPa, and a summary of the results is detailed in Table 2 below.

We observe a weaker orientation dependence in the V/N-predicted modulus (~5%) compared to the experimental values (9.7%). Part of this discrepancy may come from the in-plane (azimuthal) orientation of the indenter, which was not controlled during the experiments. The calculation of the V/N-predicted modulus assumes axisymmetric contact and therefore integrates over all azimuthal directions to obtain the average anisotropic modulus. In the experiment, the Berkovich tip has threefold symmetry and samples the in-plane deformation unevenly, causing small changes in the modulus relative to the azimuthal orientation of the tip. Because the indenter was static in our experiments and the azimuthal orientation of the samples was not controlled during loading/unloading, the coincidental alignment of the tip with a particular sample azimuth in one sample may cause a slight increase in M while causing a small decrease in another sample, inflating the observed anisotropy.

Table 2: Orientation-resolved elastic modulus and hardness of single-crystal and polycrystalline Ti 15Mo from CSM nanoindentation, before and after pile-up correction.

| Sample | $M_{\text{predicted}}$ (GPa) | $H_{\text{nominal}}$ (GPa) | $M_{\text{corrected}}$ (GPa) | $H_{\text{corrected}}$ (GPa) |
|---|---|---|---|---|
| $\langle 111 \rangle$ | 108.30 ± 1.08 | 3.29 ± 0.07 | 85.40 ± 0.94 | 3.04 ± 0.07 |
| $\langle 124 \rangle$ | 104.82 ± 1.18 | 3.35 ± 0.03 | 91.51 ± 1.46 | 3.05 ± 0.05 |
| $\langle 011 \rangle$ | 106.44 ± 1.12 | 3.39 ± 0.02 | 89.94 ± 1.62 | 3.13 ± 0.07 |
| $\langle 023 \rangle$ | 105.48 ± 1.14 | 3.54 ± 0.01 | 93.72 ± 1.57 | 3.24 ± 0.06 |
| $\langle 013 \rangle$ | 102.59 ± 1.21 | 3.59 ± 0.03 | 91.71 ± 1.39 | 3.20 ± 0.05 |
| Polycrystal | N/A | 3.38 ± 0.12 | N/A | 3.07 ± 0.12 |

The reversal of the modulus rank order is contrary to the prediction that $\langle 111 \rangle$ would be the stiffest orientation. This prediction holds for any cubic constants with A > 1 (A ~ 2.05 in our case), in uniaxial and indentation response alike [34, 36]. Additionally, previously established deformation mechanisms, such as {332}⟨113⟩ twinning (or the transformation-assisted mechanisms associated with it) and stress-induced α″ phase precipitation [38], in β-Ti alloys favor greater compliance along ⟨001⟩-side planes and greater stiffness in planes close to ⟨111⟩, further contradicting our results. Some BCC metals (Nb, Mo, V) do show ⟨111⟩ as the most compliant and ⟨001⟩ as the stiffest direction [39, 40], but this inversion originates in the d-band character and electronic properties of these alloys, causing a soft trigonal shear constant $C_{44}$ and resulting $2C_{44} < (C_{11} - C_{12})$, which leads to A < 1 [41-43]. For A ~ 2.05, this mechanism cannot produce the observed reversal, motivating a different explanation. One possible origin of the discrepancy between the measured and β-only predicted indentation-modulus ordering is an ω-related anelastic or microstructural contribution. This possibility is physically motivated because ω formation in β-Ti is associated with {111}β plane collapse and it has been shown that partially collapsed athermal ω persists in solution treated β-phase Ti-15Mo as nanoscale spheroidal particles distributed through the matrix [44, 45]. Although the amount, morphology, and variant distribution of the athermal ω-phase in the tested Ti–15Mo crystal have not yet been measured, we draw from literature on polycrystalline Ti-15Mo and other β-Ti alloys to support our discussion here.

We propose that these partially collapsed structures are mechanically accessible rather than frozen in. First-principles calculations in binary Ti–Mo place partially collapsed configurations at local energy minima along the transformation path, so a continuous range of collapse states is energetically favorable [46]. Atomic-resolution imaging has shown the completion of collapse increasing gradually from the β/ω interface into the particle interior [47], so the interface is a gradient rather than a sharp boundary and can advance or retreat through small incremental displacements without a nucleation event. Finally, it has been demonstrated that stress can drive motion along this path in both directions: the transformation proceeds forward under homogenous lattice shear [48] and backward when driven by dislocation cutting [49]. Most directly, ultrasonic measurements on a solution-treated metastable β-Ti single crystal show the athermal ω population growing and shrinking reversibly and without hysteresis under an oscillating stress field, with the accompanying dissipation attributed to local partial collapse of the {111} planes [50].

In the present study, CSM extracts stiffness from a small oscillation superimposed on the loading segment. The β/ω interfaces are demonstrably mobile at MHz frequencies, where athermal ω produces strong internal friction while leaving the elastic constants essentially unchanged [50]. At the 110 Hz oscillation frequency of our CSM tests, the extended cycle time potentially allows for the collapse to follow the applied load, so its contribution to the measured compliance would be larger, appearing here as an anelastic, low-hysteresis reduction in the apparent modulus [51]. The orientation dependence then follows from variant geometry. Four ⟨111⟩β collapse directions are available, only one of which aligns with a given ⟨111⟩ loading axis, and selection among them is governed by the stress resolved onto each [47, 52]. The present experiment does not measure frequency-dependent storage modulus, phase lag, or dissipation. We therefore interpret the CSM modulus deficit as an apparent dynamic-compliance effect whose microscopic origin remains to be tested. Based on this framework, loading near ⟨111⟩ therefore couples strongly to the aligned variant and excites the reversible response most effectively, while loading far from ⟨111⟩ couples weakly to all four, resulting in the modulus trend observed in Figure 1. This geometrical argument is consistent with the observed trend provided the underlying ω-related mechanism is sound.

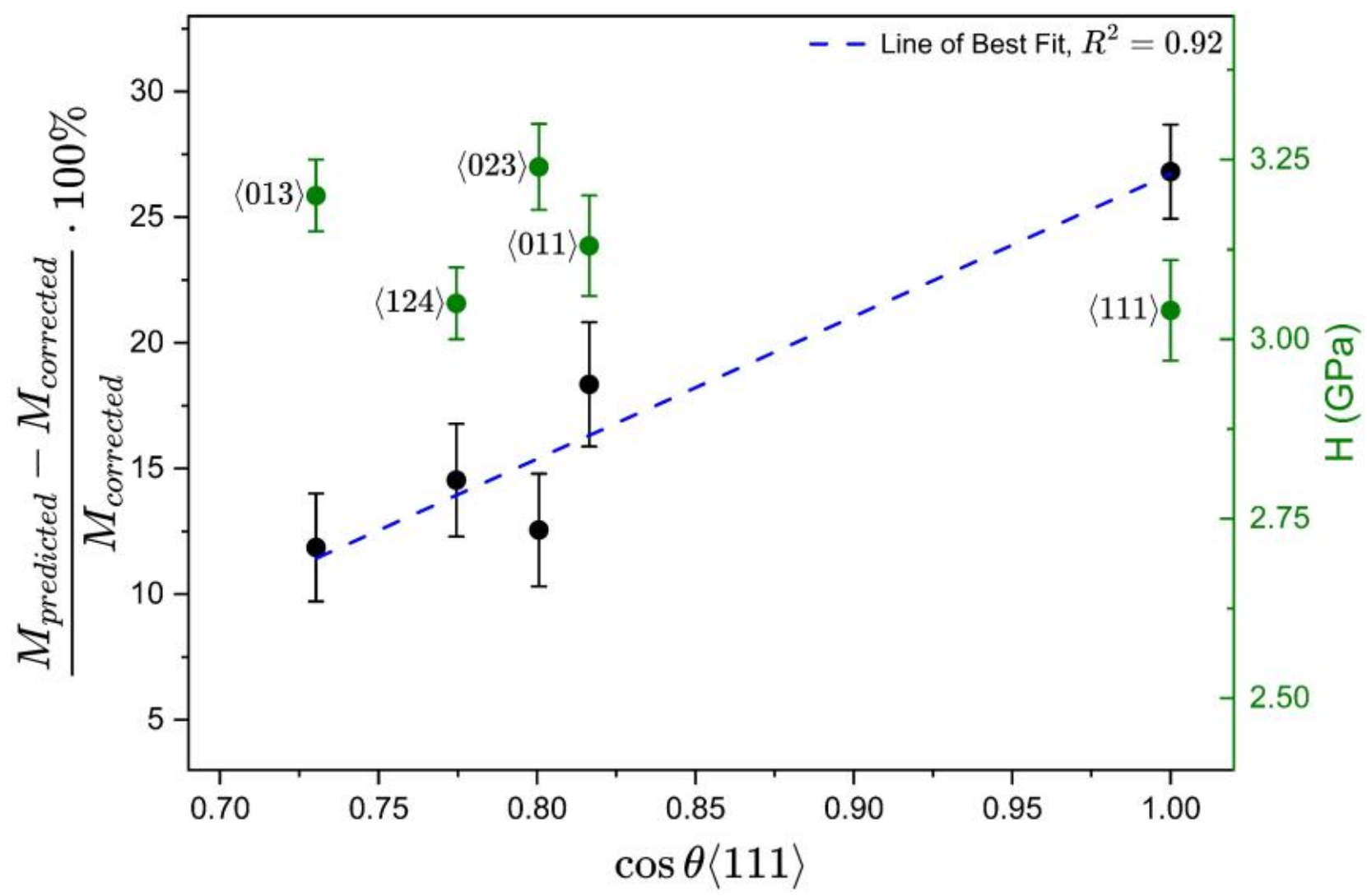

Figure 2: The percent change between the predicted and the experimental modulus vs. the cosine of the angular distance of each orientation relative to ⟨111⟩. On the right-hand side, we plot the hardness to illustrate the lack of a trend with the angular distance from ⟨111⟩.

The hardness anisotropy (6.6%) is smaller than that of the modulus, and the low hardness group (⟨111⟩, ⟨124⟩) is separated from the high hardness group (⟨011⟩, ⟨013⟩, ⟨023⟩) by approximately one standard deviation, which is at the limit of resolution. This near-isotropy is expected: hardness measures the flow stress at a representative strain of ~7–8% [53, 54]; the ω-mediated compliance that governs the elastic response or even the yield onset (section 4.3) is a small-strain, exhaustible channel and has ceased to contribute by this strain, as reflected by the lack of a trend in the hardness data points in Figure 1. The complex triaxial field beneath a Berkovich indenter, which likely activates multiple slip systems simultaneously [55], further suppressing orientation dependence.

3.3 Strain Rate Sensitivity

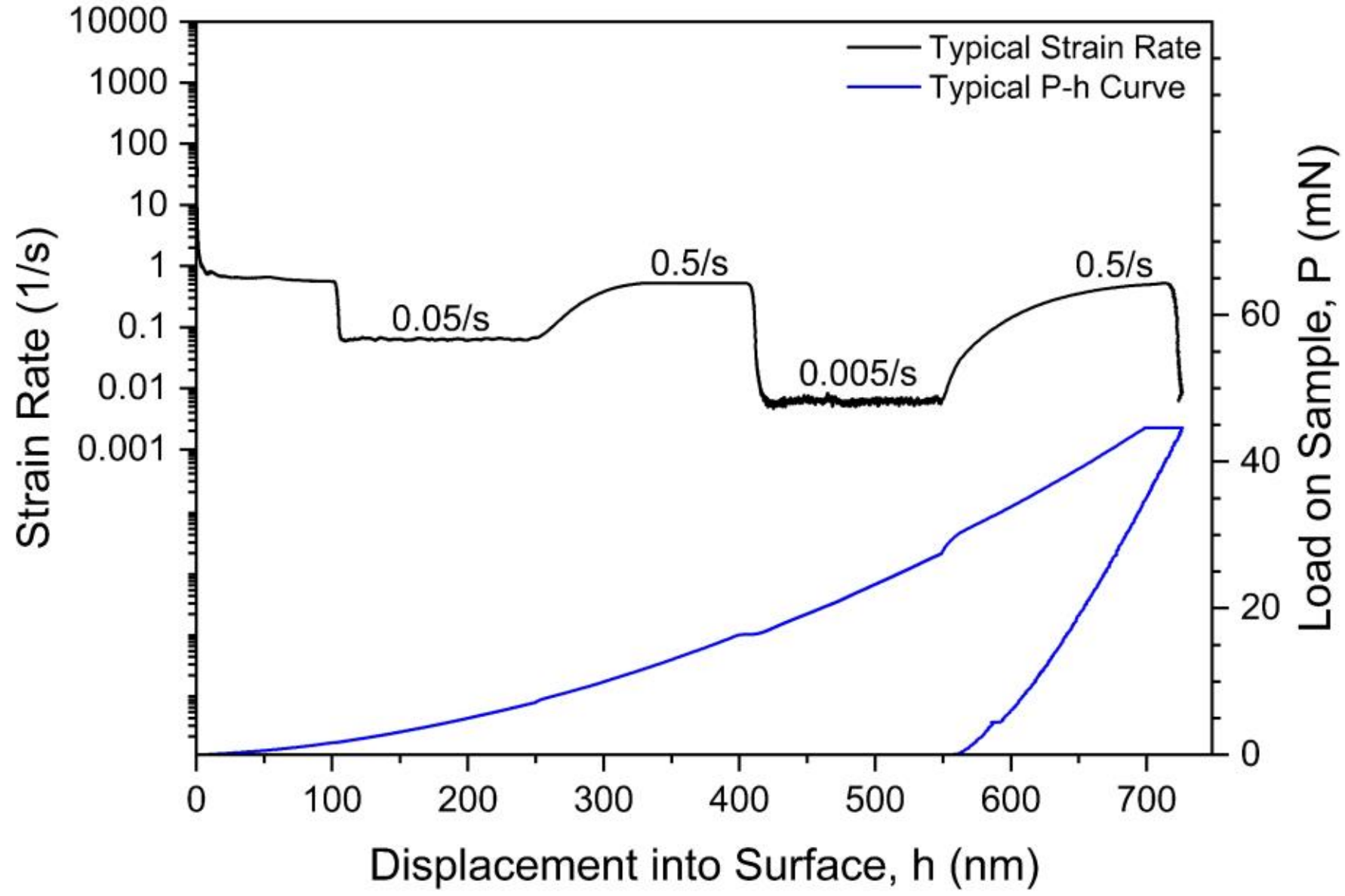

Figure 3: SRJ indentation data from one indent, showing the variation of load and strain rate with depth.

Figure 3 illustrates the typical SRJ test with data from one indent on the polycrystalline sample, where the strain rate sensitivity is calculated from the jump from 0.5/s to 0.005/s. The basis for the strain rate sensitivity is the rate-dependent hardness, as illustrated in Appendix A. The calculation of the activation volume requires a flow stress reference. We use the hardness measured at the lowest strain rate (0.005/s) to probe the quasi-static intrinsic flow stress as accurately as possible and place every orientation on consistent ground.

Table 3: Orientation-resolved strain rate sensitivity and activation volume of single-crystal and polycrystalline Ti 15Mo from SRJ nanoindentation.

| Sample | m ($10^{-3}$) | $H_{0.005/s}$ (GPa) | $M_{Taylor}$ | V* ($b^3$) |
|---|---|---|---|---|
| ⟨111⟩ | 11.87 ± 0.64 | 3.27 ± 0.04 | 3.19 | 25.62 ± 1.41 |
| ⟨124⟩ | 10.00 ± 0.71 | 3.31 ± 0.02 | 2.85 | 26.87 ± 1.91 |
| ⟨011⟩ | 9.96 ± 0.35 | 3.31 ± 0.02 | 3.18 | 30.01 ± 1.07 |
| ⟨023⟩ | 9.00 ± 0.47 | 3.47 ± 0.02 | 3.12 | 31.17 ± 1.64 |
| ⟨013⟩ | 11.99 ± 0.71 | 3.46 ± 0.02 | 2.46 | 18.46 ± 1.10 |
| Polycrystal | 12.21 ± 0.82 | 3.10 ± 0.07 | 3.00 | 24.66 ± 1.76 |

To write the apparent activation volumes in terms of the Burgers vector, we use a lattice parameter of a = 0.326 nm [13] and assume the primary deformation pathway is the ⟨111⟩ direction. The calculated values are derived from indentation hardness using an assumed hardness-to-flow-stress constraint factor C = 3. All values fall within a band of 18–31 $b^3$, placing every orientation and the polycrystal within a few tens of Burgers-vector volumes. The single-crystal orientations separate into two groups: ⟨111⟩, ⟨124⟩, ⟨013⟩, and the polycrystal cluster near 18–26 $b^3$, while ⟨011⟩ and ⟨023⟩ group near 30–31 $b^3$. The maximum spread is ~69%, between ⟨023⟩ and ⟨013⟩. The polycrystal lies at the low end of the single-crystal range and is statistically indistinguishable from ⟨111⟩ and ⟨124⟩. All the values are summarized and shown in Table 2 above. Overall, apparent activation volumes in this range in BCC metals and alloys are indicative of thermally activated plasticity, particularly temperature-dependent kink pair nucleation [56, 57]. Our values are far below the activation volume of 100–1000 $b^3$ expected for forest dislocation interactions that dominate large-grained FCC metals [56, 58, 59] and above values ranging from of 1–10 $b^3$ associated with surface dislocation nucleation [60].

Because $H$ varies by only ~6% across orientations, the activation volume anisotropy is carried almost entirely by $m$, which varies by more than 30%. The underlying mechanisms must therefore change the rate response without significantly influencing hardness. {332}⟨113⟩ twinning and stress-induced transformation could be the underlying mechanisms as they operate in Ti–Mo alloys in strongly orientation-dependent proportion [61]. Both are low-stress deformation modes: twinning operates from the onset of yield [62], and twinning-induced plasticity has been documented in Ti–15Mo at ~0.5–0.7 GPa [63, 64]. Under the ~3 GPa Berkovich contact pressure they are fully engaged at every orientation and likely do not differentiate $H$. Twinning and transformation are nearly rate-insensitive [65], so any strain carried by them is unaffected by the rate jump and lowers the measured $m$ and correspondingly inflates the $V^*$. This is consistent with the observed pattern provided that, ⟨011⟩ and ⟨023⟩ are driven by twinning/transformation and ⟨111⟩, ⟨124⟩, and ⟨013⟩ are glide-dominated.

This appears to conflict with the ⟨111⟩ elastic anomaly attributed above to stress-favored ω, but the two roles are distinguished by strain content. The β↔ω transformation is shuffle-dominated, with negligible deviatoric shape strain: it can act as a reversible anelastic compliance channel yet cannot carry a meaningful fraction of the ~8% representative plastic strain beneath the indenter. The ω-phase has been shown to switch metastable β alloys between dislocation channeling (ω present) and combined twinning/transformation (ω suppressed) [66]. If ⟨111⟩-axis loading stress-stabilizes ω, as argued in section 4.1, local twinning and transformation are disfavored and glide dominates; the elastic and plastic responses are therefore two consequences of the same ω-phase stabilization. The membership of ⟨013⟩ in the glide-like group would be paradoxical under Schmid-governed deformation, since ⟨013⟩ is the most twinning-favored orientation by resolved shear (section 4.3). However, twin selection in β-Ti is governed by the shear-shuffle path and local stability rather than by resolved shear, so a high nominal twinning Schmid factor does not guarantee twinning under the triaxial Berkovich stress state. The glide-like kinetics of ⟨013⟩ are therefore consistent with the same non-Schmid behavior that produces the non-CRSS yield ordering, and we do not attribute them to the ω-phase, which would contradict Sections 3.2 and 3.4.

This framing does not uniquely identify a plasticity controlling mechanism in ⟨013⟩; Because the indentation field is multiaxial and may involve slip, twinning, transformation, and evolving contact geometry, the reported $V^*$ values should be interpreted primarily as comparative kinetic descriptors across orientations. The picture as a whole remains a hypothesis: the ω-phase/twinning relationship is itself debated, with interfacial ω also proposed to mediate {332} twin growth [66]. Distinguishing these mechanisms requires post-mortem TEM of the ⟨013⟩ and intermediate-orientation impressions or in-situ experimentation. Furthermore, it is of interest to examine Ti-15Mo's strain rate response over a larger regime of indentation strain rates through individual indents without the constraint of the SRJ capabilities, and these are the subject of ongoing work and future publications.

## 3.4 Stress-strain Curves and Elastoplastic Behavior

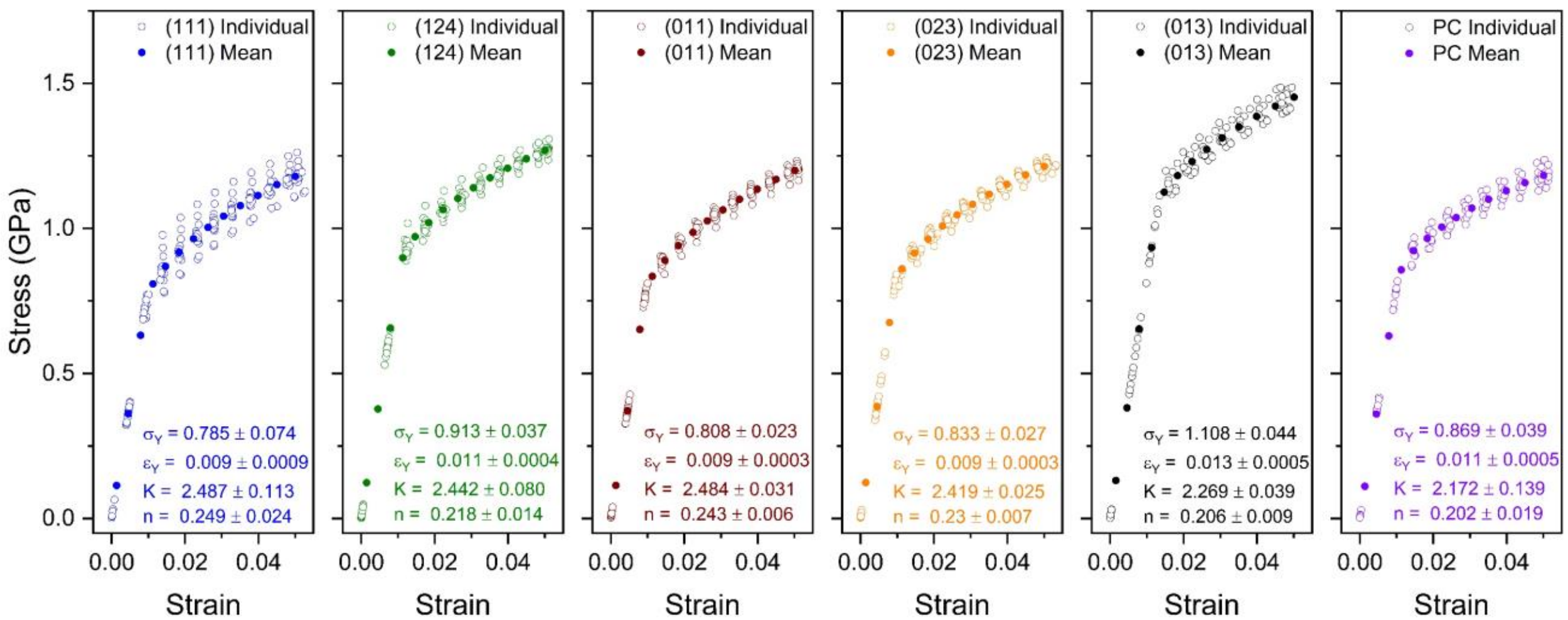

Figure 4: Stress-strain curves obtained from flat-punch indentation. The average result from each sample is plotted (filled circles) as well as the individual results from all 16 indents that were done. The yield stress, yield strain, and work-hardening exponent are also shown for each orientation. The polycrystal is labeled PC in the right-most panel.

The elastoplastic response is where orientation dependence is strongest, and the stress-strain curves are illustrated in Figure 4 above. Yield stresses ranged from 0.785 ± 0.074 GPa (⟨111⟩) to 1.108 ± 0.044 GPa (⟨013⟩), a 41.1% spread, with the polycrystal (0.869 ± 0.039 GPa) near the middle of the single-crystal range. This plasticity-onset spread far exceeds that of the modulus (~10%) and hardness (6.6%) anisotropies. The strain-hardening parameters were comparatively less orientation-dependent, with the strength coefficient (K) varying 9.6% among the single crystals and the strain-hardening exponent (n) spanning 20.9%. Notably, these trend inversely to yield stress: ⟨111⟩ combines the lowest yield with the highest K and n (greatest hardening capacity), while ⟨013⟩ pairs the highest yield with comparatively low hardening.

Table 4: Results of the ISF calculations illustrating the orientation-dependent ISFs for both twinning and slip in both the core beneath the indenter and at the rim of the indenter. $p_0$ is the maximum contact pressure under the indenter (see Appendix B).

| Sample | $p_0$ (GPa) | $m_{\text{uni,slip}}$ | $m_{\text{uni,twin}}$ | $\text{ISF}_{\text{core,slip}}$ | $\text{ISF}_{\text{core,twin}}$ | $\text{ISF}_{\text{rim,slip}}$ | $\text{ISF}_{\text{rim,twin}}$ |
|---|---|---|---|---|---|---|---|
| ⟨111⟩ | 5.64 | 0.34 | 0.21 | 0.0198 | 0.0124 | 0.279 | 0.268 |
| ⟨124⟩ | 6.38 | 0.50 | 0.46 | 0.0288 | 0.0268 | 0.273 | 0.249 |
| ⟨011⟩ | 5.85 | 0.49 | 0.33 | 0.0279 | 0.0187 | 0.266 | 0.243 |
| ⟨023⟩ | 6.04 | 0.48 | 0.42 | 0.0271 | 0.0236 | 0.266 | 0.241 |
| ⟨013⟩ | 7.25 | 0.49 | 0.49 | 0.0305 | 0.0300 | 0.271 | 0.265 |

The orientation dependence of the yield onset is large, and it is incompatible with control by a critical resolved shear stress on any deformation mode. Indentation Schmid factors (ISF) computed for the flat-punch field show two limiting behaviors: in the core of the contact and on the rim of the punch. In the contact core, on the punch axis, the stress state is axisymmetric, so its deviator is collinear with that of uniaxial compression along the loading axis. The resolved shear on any system is then τ = m ($\sigma_{zz} - \sigma_{rr}$), where m is the conventional uniaxial-compression Schmid factor (labeled $m_{\text{uni,slip}}$ in Table 4). The stress state itself is not uniaxial due to the lateral confinement by the surrounding half-space. However, because the hydrostatic component resolves to zero on every slip and twin system, this triaxial stress state does not enter the resolved shear, and the ranking of τ across orientations is identical to the uniaxial-compression Schmid ranking. This would make ⟨111⟩ the hardest orientation, carrying the lowest factors of the five orientations for both slip (0.34 vs 0.48–0.49) and compression {332}⟨113⟩ twinning (0.21 vs 0.33–0.49). The results from the ISF calculations are summarized in Table 4. Near the flat punch rim, the computed factors become nearly orientation-independent (spreads of ≤5% for slip and ≤11% for twinning), predicting a nearly isotropic onset. Even if yield is controlled by CRSS near the rim, the ISF calculations predict ⟨011⟩ and ⟨023⟩ to have the highest yield strengths, and this is not reflected in the experimental data.

Moreover, under CRSS-controlled yield, the resolved shear present on the operative system at yield must be orientation-independent, since it equals the material's CRSS. Instead, the back-calculated core resolved shear at yield across all orientations rises by ~2x for slip (0.112→0.221 GPa) and more than 3x for {332}⟨113⟩ twinning (0.070→0.217 GPa), as illustrated in Figure 5. This analysis is further supported by examining twinning specifically: in the core, the resolved shear stress is ~0.06 GPa at the ⟨111⟩ yield onset, which lies well below the {332} twin formation stress of 0.17–0.18 GPa measured in polycrystalline Ti–15Mo [67]. Based on the present core and rim ISF calculations, the measured yield-onset ordering is not reproduced by a simple orientation-independent CRSS model for the evaluated slip and {332}⟨113⟩ twinning systems. This mismatch may reflect non-Schmid effects, spatially heterogeneous indentation stresses, transformation-assisted deformation, uncertainties in the active nucleation volume, or mechanisms not represented in the model.

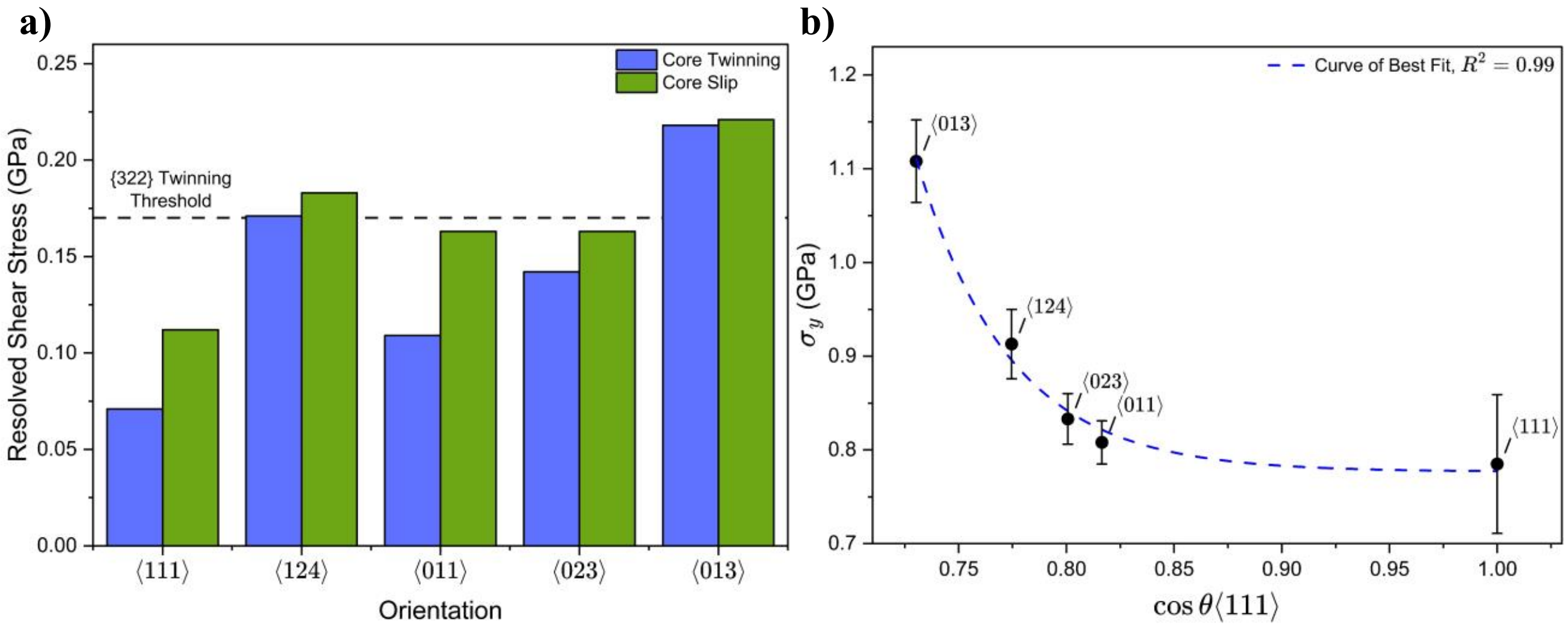

Figure 5: **a)** the resolved shear stresses at yield for each orientation (calculated from the indentation Schmid factor). Indicated on this graph is the stress threshold for {332} twinning in polycrystalline Ti-15Mo. **b)** the yield stress vs. the cosine of the angular distance of each orientation relative to ⟨111⟩.

We propose that an ω-related microstructural contribution is one possible explanation for the low apparent yield onset near ⟨111⟩ based on the work of Lai et al., who showed that ω-containing metastable β-Ti microstructures can alter the balance between dislocation channeling and TWIP/TRIP-related mechanisms [66]. This framework is consistent with the strain rate sensitivity and modulus analyses, in which ⟨111⟩ shows glide-like kinetics. If the same ω coupling that we propose softens the ⟨111⟩ yield onset also locally disfavors twinning and transformation, as reported for related β-Ti alloys [66], then glide would remain the dominant strain carrier there, tying the two observations together. Plotting the measured yield stress against the cosine of the angle relative to the ⟨111⟩ plane in the right panel of Figure 5 above, we see a roughly exponential increase in the yield strength with the increase in the distance from the ⟨111⟩ plane. Planes near ⟨001⟩, namely ⟨124⟩ and ⟨013⟩, actually approach the stress threshold for twining and slip in the core, as shown in left panel of Figure 5, further supporting the proposed mechanism of ω phase mediated yield. Note that proposing ⟨013⟩ is glide-dominated in its plastic flow response (as determined in section 4.1) despite favoring twinning at onset is not contradictory because onset reflects the initiating mode, whereas the SRJ test probes flow after strain accumulation, and {332} twinning saturates early in metastable β-Ti alloys [62].

# 4 Conclusions and Next Steps

We used EBSD guided, orientation-resolved nanoindentation to characterize the elastic and plastic anisotropy of single-crystal β Ti–15Mo alloy for surface normals near ⟨111⟩, ⟨124⟩, ⟨011⟩, ⟨023⟩, and ⟨013⟩. Modulus, hardness, strain rate sensitivity, activation volume, and yield onset each vary with orientation, but at different magnitudes, in different rank orderings, and at different strain scales. We propose these signatures are consistent with one anisotropic actor (the retained athermal ω) whose small-strain, exhaustible compliance channel leaves a different fingerprint at each strain scale. The main findings are as follows:

- The apparent indentation modulus does not follow the β-only elastic prediction. Using published single-crystal elastic constants and anisotropic indentation theory, orientations near ⟨111⟩ are expected to be the stiffest. Experimentally, however, the lowest apparent indentation modulus is measured near ⟨111⟩, indicating a reversal of the predicted ordering. This discrepancy suggests that the measured dynamic indentation response contains a contribution beyond homogeneous β-phase elasticity.
- The pile-up corrected hardness  varies by only 6.6%, with group separation at the 1σ level, and sorts with neither the yield ordering nor the rate-sensitivity groups. The Berkovich indenter measures hardness at a representative strain of 7–8% where any yield anisotropy caused by the ω phase is no longer dominant, making the hardness much less anisotropic than the small-strain elastic response and the onset of plasticity. Furthermore, the complex triaxial stress state induced by the indenter activates multiple slip systems at once. These factors contribute to the weak orientation dependent anisotropy observed in the hardness.
- Activation volumes of 18–31 $b^3$ with $m \approx 10^{-2}$ are the consistent with the BCC kink pair nucleation mechanism, well below forest-interaction and above surface-nucleation volumes. The activation volumes group the orientations in two: glide-dominated (⟨111⟩, ⟨013⟩, ⟨124⟩) and twin/transformation-participating (⟨011⟩, ⟨023⟩). Although ⟨013⟩ is the most twin-favored orientation per our Schmid analysis for the flat punch indenter, we propose that the {332} twinning mechanism is largely exhausted at the high strains associated with plastic flow and SRJ indentation. Although these values provide useful comparative kinetic descriptors across orientations, they do not by themselves uniquely identify the controlling deformation mechanism under the multiaxial indentation stress state.
- The flat-punch yield-onset ordering is not reproduced by the simple indentation-Schmid-factor/constant-CRSS models evaluated for slip and {332}⟨113⟩ twinning. The yield onset varies by 41% (⟨111⟩ softest), incompatible with critical-resolved-shear-stress control: the resolved shear at yield varies ~2× and ~3× across orientations for slip and {332} twinning, respectively, where yield by CRSS would reflect isotropic resolved shear stresses. The ISFs also tell a different story: the opposite ordering (⟨111⟩ hardest) in the core and the rim predicts near-isotropy; neither reproduces the observed spread. These results imply a non-CRSS onset mechanism, which is proposed to be ω-mediated channeling.

Overall, the results demonstrate that the local mechanical behavior of single-crystal Ti–15Mo is strongly orientation dependent and cannot be fully rationalized by β-phase elasticity and simple resolved-shear-stress arguments alone. Direct verification of the proposed ω-related interpretation will require site-specific diffraction or TEM, ideally combined with pre- and post-indentation characterization and, if possible, frequency-dependent measurements of stiffness and dissipation.

**Acknowledgements**

K.A. thanks the Yale Dean's Office for partially funding this research through the Science, Technology, and Research Scholars II (STARS II) fellowship. J.Š. acknowledges financial support of the Ministry of Education, Youth and Sports of the Czech Republic (MEYS) through the Operational Programme Johannes Amos Comenius, project FerrMion (CZ.02.01.01/0022_008/0004591), co-funded by the European Union, and of the Czech Science Foundation through project 24-14578L. Single crystal growth was performed in MGML (mgml.eu), which is supported within the program of Czech Research Infrastructures, project no. LM2023065. A.D thanks collaborators at KLA, Oak Ridge National Laboratory, and Idaho National Laboratory for their support. Nanoindentation, atomic force microscopy and electron backscatter diffraction measurements were performed at the Yale Mechanical and Thermal Characterization Core, Yale Institute of Quantum Engineering and the Yale Earth Materials Characterization Center, respectively. The authors thank the facility staff for training and technical assistance. The authors declare no competing financial interests or other conflicts of interest.

**Appendix A: Nanoindentation Methods and Pile-up Correction**

CSM Indentation Pile-up Correction

Using the reduced modulus $E_{\mathrm{ru}}$ obtained from CSM indentation and after pile-up correction (see below), and the properties of the diamond indenter, we can calculate the true modulus of the sample using:

$$\frac{1}{E_{\mathrm{ru}}} = \frac{1-\nu_i^2}{E_i} + \frac{1-\nu^2}{E} \tag{7.1.2}$$

The equation above assumes isotropy and is only applied to the polycrystal. Furthermore, this equation requires knowledge of the sample's Poisson's ratio. For the polycrystal, the Poisson's ratio is estimated using the method outlined by Zorzi and Perottoni [68] in conjunction with equation (7.1.2) to estimate the Poisson's ratio and Young's modulus:

$$H_v = 2\left[\frac{9E}{8}\frac{(1 - 2\nu)^2}{(1 + \nu)^3}\right]^{0.585} - 3 \tag{7.1.3}$$

where $H_v$ is the Vicker's hardness. We can estimate the Vickers hardness by applying a scaling factor to the indentation hardness (in MPa) obtained from the Berkovich indenter: $H_v = 0.0925\, H_{IT}$. This yields $\nu = 0.3$, which is in excellent agreement with accepted values (0.3-0.35) for annealed β-Ti alloys [17]. It should also be noted that varying the Poisson's ratio over a range of 0.3 to 0.35 causes a minimal ~3.6% spread in the modulus for our polycrystalline sample.

For the anisotropic orientations, we compute the anisotropic modulus M:

$$\frac{1}{E_{\mathrm{ru}}} = \frac{1-\nu_i^2}{E_i} + \frac{1}{\mathrm{M}} \tag{7.1.4}$$

The pile-up correction was performed by adopting the method of Qui, Y. H. et al. [30]. In practice, the majority of nanoindentation experiments result in asymmetric pile-ups, as governed by the slip systems, deformation symmetry, azimuthal angle of the indenter, and more. To account for this asymmetry, the three pile-up heights are averaged, and a constant average pile-up height is used in the procedure above. We propagated uncertainties using the Monte Carlo method, drawing $N = 10^5$samples of each input (the average pile-up height, the nominal hardness, etc.) from their measured distributions, evaluating the full correction model on each draw, and reporting the resulting mean and standard uncertainty for the corrected hardness and reduced modulus. The figure below illustrates an AFM map of the typical CSM indent and the resulting pile-up profiles.

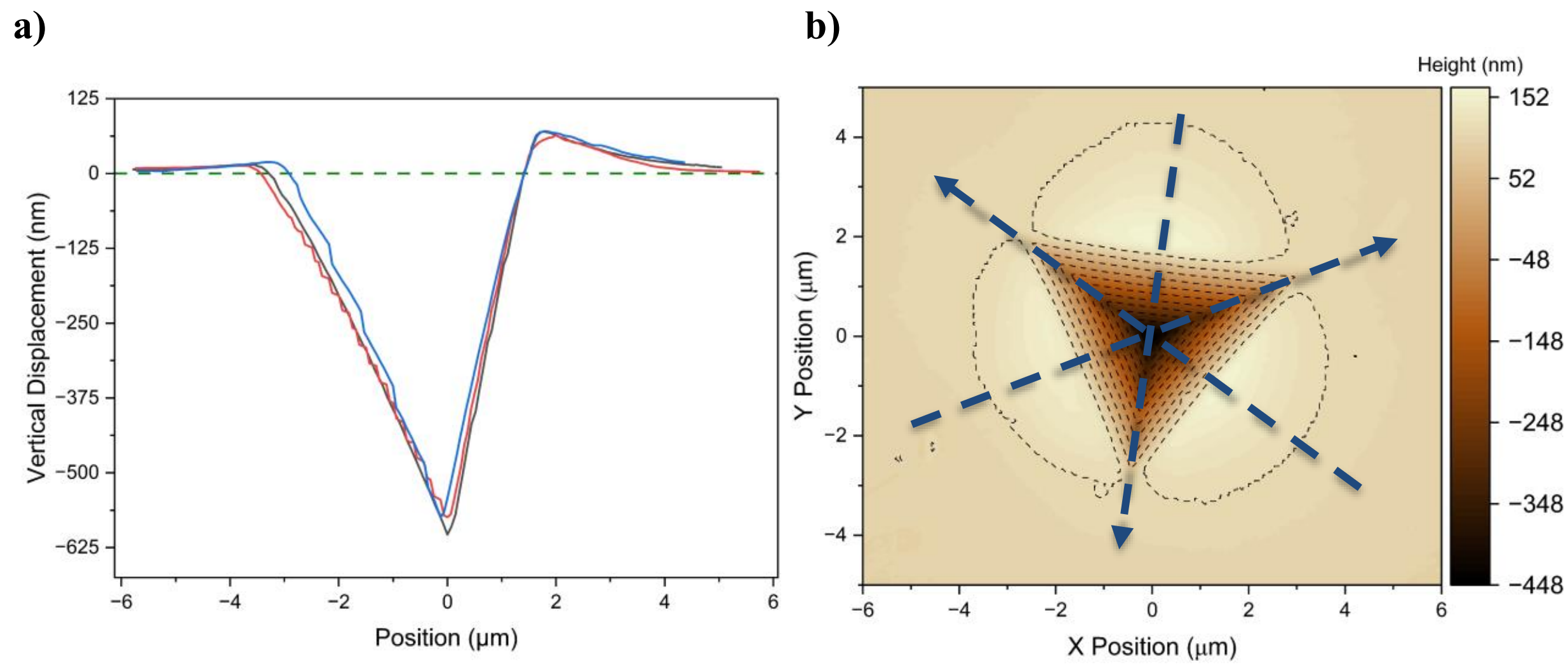

Figure 6: **a)** The AFM profiles of a representative CSM indent, where all three pile-up lobes are overplotted. **b)** an AFM contour map of the CSM indent, where the dashed arrows indicate the lines along which the line scans were performed to obtain a two dimensional pile-up profile. The pile-up lobes are outlined by the outer-most contour lines and the color map is zeroed on the sample surface.

## Strain Rate Sensitivity and Activation Volume

Strain rate sensitivity is measured by characterizing the rate-dependent change in the material's resistance to deformation (i.e. hardness in the case of indentation) or flow stress (uniaxial testing) [69-71]:

$$m = \frac{\partial(\ln \sigma)}{\partial(\ln \dot{\varepsilon}_{uniaxial})} = \frac{\partial(\ln H)}{\partial(\ln \dot{\varepsilon}_{indentation})} \tag{7.2.1}$$

Note that the indentation strain rate is the indentation equivalent to the uniaxial strain rate and is approximated as follows [72]:

$$\varepsilon = \frac{\dot{h}}{h} = \frac{1}{2}\left(\frac{\dot{P}}{P} - \frac{\dot{H}}{H}\right) \approx \frac{\dot{P}}{2P} \tag{7.2.2}$$

where the final approximation emanates from the assumption that hardness remains constant with indentation depth. We also calculate the apparent activation volume [73]:

$$V* = C\sqrt{3}\frac{kT}{mH} \quad (7.2.3)$$

where k is Boltzmann's constant, T is the test temperature, and C is assumed to be a rate-independent constant relating hardness to yield strength. Following Tabor's relation, we adopt C ≈ 3 [53, 74] for the polycrystal. For the single crystals, we adopt C = $M_{Taylor}$, where $M_{Taylor}$ is the anisotropic Taylor factor is computed for each sample using the MTEX Toolbox (MATLAB R2024b).

**Appendix B: The Flat Punch Indentation Schmid Factor**

Here, we outline the method used to compute indentation Schmid factors (ISFs) for the flat-punch indentation of each orientation in this study. The analysis tests whether the measured orientation dependence of the yield onset can be attributed to resolved shear stresses on the available slip and twinning systems. We generalize the Hertzian indentation Schmid factor method of Li et al, 2011 [32] to the flat-punch geometry, modelling the punch with its true, as-manufactured rounded edge so that the peak contact pressure is finite.

The Flat Punch Contact Problem

A perfectly sharp rigid cylinder of radius b pressed to depth δ into an elastic half-space of reduced modulus $E_{ru}$ produces a contact pressure profile given by [75]:

$$p(r) = \frac{P}{2\pi b\sqrt{b^2 - r^2}}, \quad P = 2E_{ru}b\delta \quad (8.1.1)$$

The pressure is minimum at the center and diverges at the rim (r → b), the signature of the infinitely sharp corner. In reality, the flat punch indenter is not a perfect cylinder with an infinitely sharp corner. The punch has a flat circular face of radius b = 1.08 µm (2b = 2.16 µm) and a cylindrical body of radius R = 1.125 µm (2R = 2.25 µm), joined by a rounded shoulder. Assuming the punch possesses a smooth fillet joining the two radii, a circular arc tangent to both the flat face and the cylindrical wall must have radius:

$$R_e = R - b = \frac{2.25 - 2.16}{2}\ \mu\text{m} = 45\ \text{nm} \quad (8.1.2)$$

To account for this geometry, we consider a general axisymmetric profile f(r) and define the Sneddon transform:

$$g(t) = t\int_0^t \frac{f'(r)}{\sqrt{t^2 - r^2}}\,dr \quad (8.1.3)$$

which is the indentation depth at which the contact radius reaches a value t (an Abel transform of the profile). Mossakovskii's method [76] views the loading as a continuous stack of ideal-flat-punch increments of growing radius: as the punch advances by dδ with the contact edge at radius b, the incremental solution is that of an ideal flat punch of radius b. Summing the increments gives

$$\delta = g(a) \quad (8.1.4)$$

$$P = \int_b^a 2E_{ru}t\,g'(t)\,dt \quad (8.1.5)$$

$$p(r) = \frac{E_{ru}}{\pi} \int_{max(r,b)}^{a} \frac{g'(t)}{\sqrt{t^2 - r^2}} \, dt \tag{8.1.6}$$

where a is the final contact radius. The pressure at r superposes edge-of-flat-punch contributions from all increments whose radius has passed r; each carries the inverse-square-root form (8.1.1), but smeared over t ∈ [b, a]. The smearing renders the rim pressure finite.

## 4.1 The Solution for the Rounded Punch

For our rounded flat punch, we define the shoulder profile as the height above the flat punch face versus radius r:

$$f(r) = 0 \ (r \leq b) \tag{8.2.1}$$

$$f(r) = R_e - \sqrt{R_e^2 - (r - b)^2} \approx \frac{(r - b)^2}{2R_e} \ (r > b) \tag{8.2.2}$$

The parabolic form is the leading-order approximation to the arc; comparing exact arc to this parabolic approximation, we find the two agree to better than 1% over the operating range (see the calculation of the projected rim contact half-width). With this approximation, we compute:

$$g(t) = \frac{t}{R_e}\left[\sqrt{t^2 - b^2} - b \cdot arccos\left(\frac{b}{t}\right)\right] \tag{8.2.3}$$

$$g'(t) = \frac{1}{R_e}\left[2\sqrt{t^2 - b^2} - b \cdot arccos\left(\frac{b}{t}\right)\right] \tag{8.2.4}$$

Note that taking the limit of b → 0 recovers the Hertzian contact solution exactly. The contact radius (a) at yield is calculated using the mean contact pressure (i.e. the reported yield stress) and the relationship between the contact pressure and contact radius:

$$p_{mean}(a) = \frac{P(a)}{\pi {R_{body}}^2} = \frac{2E_{ru}}{\pi \mathrm{R}^2} \int_b^a t \, g'(t) \, dt \tag{8.2.5}$$

where P(a) is the load at yield. The equation above is inverted given a yield stress (and therefore a load at yield) to determine the required contact radius to realize that stress. Given the contact area at yield and the punch face diameter, we calculate the half-width of the projected contact at the rim using w = a − b, resulting in a range of 8 to 10 nm. This half-width represents the projected contact radius of the rounded edge beyond the punch face. Furthermore, because w = a − b ≪ b, the sample surface does not rise far above the flat punch face (w ~ 0.8% of the punch face radius), indicating that approximating the rounded rim as a parabola is reasonable.

## The Internal Stress Field and Boussinesq Superposition

Linear elasticity permits superposition: the stress field beneath an arbitrary surface pressure distribution p(r) is the integral of point-force (Boussinesq) fields weighted by the local pressure,

$$\sigma(\mathbf{x}) = \int \sigma_{Bouss}(\mathbf{x} - \boldsymbol{\xi}) \, p(\boldsymbol{\xi}) \, dA \tag{8.3.1}$$

The Boussinesq solution for a normal point force F at the surface origin gives, at in-plane distance ρ and depth z (where $r^2 = \rho^2 + z^2$) [77]:

$$\sigma_z = -\frac{3F}{2\pi}\frac{z^3}{r^5}, \quad \tau_{\rho z} = -\frac{3F}{2\pi}\frac{\rho z^2}{r^5} \tag{8.3.2}$$

$$\sigma_\rho = \frac{F}{2\pi}\left[\frac{1 - 2\nu}{\rho^2}\left(1 - \frac{z}{r}\right) - \frac{3\rho^2 z}{r^5}\right] \tag{8.3.3}$$

$$\sigma_\theta = -\frac{F(1 - 2\nu)}{2\pi}\left[\frac{1}{\rho^2}\left(1 - \frac{z}{r}\right) - \frac{z}{r^3}\right] \tag{8.3.4}$$

Note that $\sigma_z$ and $\tau_{\rho z}$ are independent of the Poisson's ratio (fixed by equilibrium), while the in-plane normal stresses carry the Poisson response through the $(1 - 2\nu)$ terms.

*The Flat Punch Stress Field*

At any interior point the stress state is the symmetric second-rank tensor σ. On the axis of symmetry beneath the punch center, axisymmetry forbids shear components and equates the two lateral normal stresses, so the tensor reduces to a hydrostatic part plus a single uniaxial term:

$$\sigma = \sigma_{rr}\,\mathbf{I} + s(z)\,\hat{z} \otimes \hat{z}, \quad s(z) = \sigma_{zz} - \sigma_{rr} < 0 \tag{8.4.1}$$

For a deformation system with unit slip-plane normal **n** and unit shear direction **d** (with **n** ⊥ **d**), the resolved shear stress driving that system is the double projection:

$$\tau = \mathbf{n} \cdot \sigma \cdot \mathbf{d} = \sum n_i \sigma_{ij} d_j \tag{8.4.2}$$

A key consequence follows immediately: a hydrostatic stress $\sigma = -p\,\mathbf{I}$ induces zero shear on every system because $\tau = -p\,(\mathbf{n}\cdot\mathbf{d}) = 0$ whenever **n** ⊥ **d**. Only the deviatoric (non-hydrostatic) part of the stress can shear a crystallographic plane, which is simply the uniaxial term in this case. As a result, the resolved shear on any system becomes

$$\tau = s(z)\,(\mathbf{n} \cdot \hat{z})(\mathbf{d} \cdot \hat{z}) = s(z) \cdot m \tag{8.4.3}$$

where m is the ordinary uniaxial Schmid factor for loading along the surface normal. Because $s(z) < 0$, the on-axis deviatoric state is uniaxial *compression*, which selects the compression sense of the polar twinning systems. To compute s(z), we first determine the on-axis stresses by evaluating the superposition integral (8.3.1), substituting the flat-punch pressure distribution (8.1.6) and integrating the point-force components (8.3.2) to (8.3.4):

$$\frac{\sigma_{zz}(z)}{p_{mean}} = -\frac{1}{2}\frac{a^2(a^2 + 3z^2)}{(a^2 + z^2)^2} \tag{8.4.4}$$

$$\frac{tr\,\sigma(z)}{p_{mean}} = -(1 + \nu)\,\frac{a^2}{a^2 + z^2} \tag{8.4.5}$$

$$\sigma_{rr} = \frac{tr\,\sigma - \sigma_{zz}}{2} \tag{8.4.6}$$

The quantities above have been normalized by the mean contact pressure to render the final calculated s(z) value load-independent. s(z) attains its maximum magnitude at depth $z \approx 0.94a$, and we define $s_{max}(\nu) = max_z|s(z)|$, which depends only on ν, ranging from 0.421 at $\nu = 0.32$ to 0.403 at $\nu = 0.39$ (a higher Poisson ratio makes the state more hydrostatic and slightly reduces the deviatoric drive). Physically, $s_{max}$ is a constraint efficiency: in a uniaxial test the lateral surfaces are free, so $\sigma_{rr} = 0$ and $s = \sigma$ and the applied stress is transmitted in full to the deviatoric drive; the familiar $\tau = \sigma m$ is the special case of $\tau = sm$ with $s = \sigma$. Beneath an indenter the loaded volume is laterally confined by the surrounding material, $\sigma_{rr} \neq 0$, and only the fraction $s_{max}(\nu) \approx 0.40$–$0.42$ of the mean contact pressure survives as deviatoric drive, the remainder being hydrostatic and unable to shear any crystallographic plane. The core resolved shear is therefore:

$$\tau_{core} = s_{max}(\nu) \cdot p_{mean} \cdot m \tag{8.4.7}$$

4.2 Computing the Uniaxial Compression Schmid Factor

Projecting a uniaxial stress of magnitude σ along a loading direction **L**, for which σ = σ ($\hat{\mathbf{L}} \otimes \hat{\mathbf{L}}$), gives τ = σ m where:

$$m = (\hat{\mathbf{L}} \cdot \hat{\mathbf{n}})(\hat{\mathbf{L}} \cdot \hat{\mathbf{d}}) = \cos\varphi \cos\lambda \tag{8.5.1}$$

where φ is the angle between the loading axis and the slip-plane normal and λ that between the loading axis and the shear direction. The factor reported for each mode family is the maximum over that family, but the maximization criterion differs between slip and twinning.

Slip is bidirectional: a dislocation may glide along either ±**d**, so the sign of m is immaterial, and the magnitude is taken over all 24 systems of the {110}⟨111⟩ and {112}⟨111⟩ families:

$$m_{slip} = max_{24\,systems} \mid m \mid \tag{8.5.2}$$

Twinning is unidirectional (polar) [78]: a variant shears in one sense only, so a variant of the opposite sense cannot twin however large |m| may be. Because the on-axis deviatoric state is compressive, the compression-sense variants are the active ones, and the maximization retains only that sense over the twelve {332}⟨113⟩ variants:

$$m_{twin} = max_{12\,variants}\,(-m), \text{retaining only } -m > 0 \tag{8.5.3}$$

Under tension, ⟨111⟩ would be the most twin-favored of the six axes (m = 0.429), whereas under the compression it is the least favored (m = 0.171).

4.3 Computing the ISF

The only difference between the ISF and the uniaxial compression m in this formalism is that the former is computed relative to the peak contact pressure, following the definition proposed by Li et al. (2011) [32], while the latter is computed relative to the mean contact pressure (i.e. the yield stress).

Each slip and twin system is rotated into the laboratory frame by the orientation matrix that places the loading direction ⟨uvw⟩ along the surface normal; the in-plane azimuth ψ is scanned over 0–360° in 5° steps because the experimental azimuth is unconstrained. The resolved shear at each field point is computed using the steps outlined above. The indentation Schmid factor of each mode is evaluated separately in the core and in the rim region:

$$ISF_{mode} = max_{x,\psi,sys}\left[\frac{\tau(\mathbf{x})}{p_0}\right] \tag{8.6.1}$$

For the core, the core ISF is simply:

$$ISF_{core} = \frac{s_{max}(\nu)\, p_{mean}\, m}{p_0} = m \cdot \frac{s_{max}(\nu)\, p_{mean}}{p_0} \tag{8.6.2}$$

The core ISF preserves the uniaxial Schmid ordering exactly. It is reported only to place the core on the same normalization as the rim; the physically meaningful core quantities are the uniaxial factors m.

For the rim, the stress state is fully triaxial, retains non-zero off-diagonal components in the stress tensor, and varies with both radial position and depth, so the resolved shear must be obtained numerically. The interior field is assembled by Boussinesq superposition (section 8.3) over the rounded-punch pressure distribution p(r) (section 8.1, 8.2), which includes the finite rim peak. The field is then evaluated on a local grid spanning the near-rim region, extending radially from inside the flat face to beyond the contact edge and into the sample to several times the shoulder contact width (w), with coordinates scaled by w. At every grid point the full stress tensor is contracted with every slip and twin system for every azimuth, and the maximum of $\tau/p_0$ over grid points, systems and azimuths is the rim ISF.